\documentclass[apj]{emulateapj}
\slugcomment{{\sc Accepted for publication in the ApJ}}
\def\xsrc{4U~0142+61~}
\def\xsrcnos{4U~0142+61}

\def\rxs{1RXS~J170849.0-400910~}
\def\esrcs{1E~1841-045~}
\def\esrc{1E~1841-045}
\def\igl{INTEGRAL~}
\def\iglnos{INTEGRAL}
\begin{document} 
\vspace{0.8 in} 

\title{Search for High-Energy Gamma-ray Emission from an Anomalous X-ray Pulsar, \xsrc} 

\author{Sinem \c{S}a\c{s}maz Mu\c{s}\altaffilmark{1},
Ersin {G\"o\u{g}\"u\c{s}}\altaffilmark{1}
}
 
\altaffiltext{1}{Sabanc\i~University, FENS, Orhanl\i - Tuzla, Istanbul 34956 Turkey}


\begin{abstract} 

Until 2004, anomalous X-ray pulsars (AXPs) were known as strong emitters of soft X-rays only ($<$ 10 keV). The discovery of hard X-ray component from AXPs provided important insight about their emission properties while it posed a serious challenge to explain its origin. The physical mechanism of the hard emission component has still not been fully resolved. We investigate the high-energy gamma-ray properties of the brightest AXP, \xsrc using data collected with the Large Area Telescope on board \textit{Fermi Gamma-ray Space Telescope} to establish the spectral behavior of the source on a very broad energy span and search for pulsed emission.
Here, we present our results of detailed search for the persistent and pulsed high-energy gamma-ray emission from 4U 0142+61 which result in no significant detection. However, we obtain upper limits to the persistent high-energy gamma-ray emission flux which helps us to constrain existing physical models. 

\end{abstract} 


\section{Introduction} 

Anomalous X-ray pulsars (AXPs) have been intriguing sources since their discovery in 
early 1980s \citep{fg81}. They are bright X-ray sources with X-ray luminosities (below 10 keV) in 
the range of $10^{33}$-$10^{36}$ erg s$^{-1}$. 
They spin rather slowly and their spin periods clustered in a narrow range of 2-12 s. 
The lack of any evidence for binary nature (e.g., Doppler modulation in their long term pulse periods) eliminates the possibility that 
they are accreting matter from a donor. Their spin down rates are relatively large, ranging between $10^{-10}$ and $10^{-12}$ s s$^{-1}$.
Their rotational energy loss is insufficient by orders of magnitudes to provide the observed X-ray luminosities. 
AXPs are commonly regarded as young isolated neutron stars that are powered by their extremely strong magnetic fields, 
\textit{B}$\gtrsim$ $10^{14}$ G \citep{dt92}. Such strong magnetic fields can efficiently slow these young systems down via magnetic breaking 
and provide energy for the emitted X-rays via diffusion of evolving magnetic field \citep{td96}. A detailed review on AXPs can be found in \citet{mereghetti08} and \citet{wt06}. 

A major observational development in AXP studies was the discovery of hard X-ray emission from AXPs \esrcs \citep{kuiper04}, \xsrc \citep{dH04} and \rxs \citep{revn04}. \esrc, located at the center of supernova remnant (SNR) Kes 73, is the first AXP from which non-thermal pulsed hard X-ray/soft gamma-ray emission was discovered \citep{kuiper04}. The pulsed nature of this emission eliminates the possibility of its 
SNR origin. Spectral studies of this source using \igl observations in the
20$-$300 keV band revealed a power$-$law shape with an index of 1.39 $\pm$ 0.05 but no 
evidence of spectral break \citep{kuiper06}. The hard X-ray emission from \rxs was
discovered in the Galactic Plane Survey observations with \igl \citep{revn04} 
and shown to be pulsed as well (\citet{kuiper06}, \citet{dH08}). 
Finally, hard emission component from \xsrc was discovered during \igl observations of the 
Cassiopeia region \citep{dH04}. Detailed studies with 2.37 Ms INTEGRAL 
observations showed a power law spectrum up to about 230 keV with an index
of 0.93 $\pm$ 0.06 \citep{dH08}. Based on the logparabolic function fit to
the \igl SPI and ISGRI observations, they estimate a peak energy of the spectral energy 
distribution to be $\sim$228 keV \citep{dH08}, and 20$-1$50 keV flux as 
(8.97 $\pm$ 0.86) x $10^{-11}$ erg cm$^{-2}$ s$^{-1}$. It is important to note 
that the energy emitted above 15 keV is comparable or larger than that in the soft X-ray band 
(that is, below 10 keV, see e.g., \citet{dH08}).

The physical nature of the hard X-ray / soft gamma-ray emission is still not well 
understood. \citet{bt07} proposed that the hard 
X-ray emission could originate from a plasma corona around the magnetar.
They suggest that such a corona around magnetars can be formed via
starquakes that could shear the neutron star crust and its external magnetic
field. \citet{hh05a,hh05b} proposed the fast-mode break down model
in which an optically thick fireball produced by the magnetohydrodynamics waves 
created near the surface of the neutron star. 
Further they suggest that if fast modes are not strong enough to yield an optically thick fireball, the produced 
non-thermal emission would be sufficient to explain the observed high-energy emission from soft gamma repeaters and AXPs. 
Another attempt to explain the hard X-ray emission is by \citet{bh07}.
They suggest that the upscattered surface thermal X-ray photons as the source of observed 
non-thermal hard X-ray emission from AXPs. 

In order to understand the nature of the hard emission component of AXPs, it is crucial 
to establish their spectral shapes on a wide range in the energy domain. 
In particular, it is important to determine where their spectral energy distribution peak. 
Thanks to the Large Area Telescope (LAT) on board \textit{Fermi Gamma-ray Space Telescope (Fermi)}, we 
are now able to investigate high-energy behavior of these sources with an unprecedented data quality. 

In this paper, we performed detailed search for persistent and pulsed high-energy gamma-ray emission
from \xsrc using \textit{Fermi}/LAT observations. We also employed contemporaneous \textit{Rossi X-ray Timing Explorer (RXTE)} observations to obtain the spin ephemeris of the source. In Section 2, we describe the observations used and details of data analysis. We present our results in Section 3 and discuss their implications in Section 4.

\section{Observations and Data Analysis}
 
\subsection{{\it Fermi}/LAT}

LAT is one of the two instruments on board \textit{Fermi} operating in 
the energy band of 20 MeV$-$300 GeV. It is a pair conversion telescope with a high-resolution 
silicon tracker, calorimeter, anticoincidence detector, programmable trigger and data acquisition 
system \citep{atwood09}. Since 2008 August 4, the LAT has been operating as an all sky monitor 
in high-energy gamma-rays, covering the full sky in approximately every 3 hr.

\begin{figure*}
\begin{center}
\includegraphics[width=0.8\textwidth]{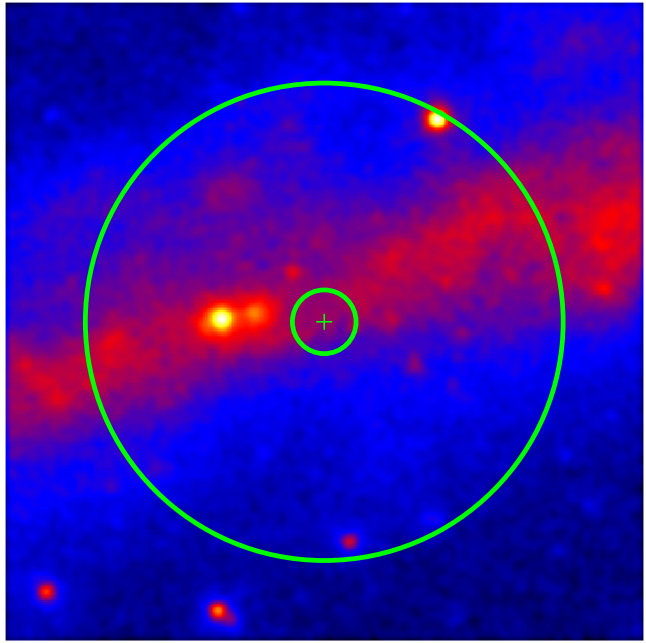}
\label{fig:countmap}
\end{center}
\caption{Smoothed \textit{Fermi}-LAT count map around AXP \xsrc in the 0.2$-$100 GeV energy range. The large circle and small circle show the $15^{\circ}$ and $2^{\circ}$ radius extraction regions, respectively. The plus sign indicates the position of AXP 4U 0142+61.}
\end{figure*}

We accumulated the LAT data within a $15^{\circ}$ radius\footnotemark \footnotetext{As suggested by the \textit{Fermi}/LAT instrument team, \\
http://fermi.gsfc.nasa.gov/ssc/data/analysis/documentation/Cicerone/Cicerone$\mathunderscore$Likelihood/Choose$\mathunderscore$Data.html} centered at \xsrcnos, collected from 2008 August 4 to 2010 April 29 with an exposure time of $\sim$31.7 Ms. We also performed spectral analysis using a $2^{\circ}$ radius region around the source in order to completely avoid contamination from the nearby bright sources . 
We performed our unbinned likelihood analysis 
\footnotemark \footnotetext{http://fermi.gsfc.nasa.gov/ssc/analysis/scitools/likelihood$\mathunderscore$tutorial.html} using ScienceTools v9r15p2 with P6$\mathunderscore$V3$\mathunderscore$DIFFUSE set as the instrumental response.  
In the event selection process, we set the maximum zenith angle to $105^{\circ}$ in order to 
eliminate background gamma-rays due to the Earth limb.
All time intervals where the zenith cut intersects the region were excluded. 
In Figure \ref{fig:countmap}, we present the count map image of the $40^{\circ}$ radius 
region centered at \xsrc.
The diffuse gamma-ray emission from the Milky Way was modeled with the latest model,  
gll$\mathunderscore$iem$\mathunderscore$v02.fit. We also used isotropic$\mathunderscore$iem$\mathunderscore$v02.txt to account for the extragalactic 
isotropic diffuse emission and residual instrumental background.
The spectral fits and flux calculations were done with the python version of gtlike, pyLikelihood.

For timing analysis, photon arrival times of all events within the $2^{\circ}$ extraction region around \xsrc were converted to that at the solar system barycenter using gtbary of ScienceTools.

\subsection{{\it RXTE}}
To search for pulsed high-energy gamma-ray emission from \xsrcnos, we obtained the precise spin ephemeris of the source using contemporaneous \textit{RXTE} observations. On board \textit{RXTE}, there are three scientific payloads: the Proportional Counter Array (PCA) that is sensitive to photon energies between 2$-$60 keV, the High Energy X-ray Timing Experiment, sensitive to photons in the 15$-$250 keV photons and the All Sky Monitor. We have employed only the PCA observations to achieve our goal.

\xsrc has been monitored with the \textit{RXTE} periodically for the last $\sim$8 years with pointings almost uniformly spaced, usually by about two weeks. We have selected 53 \textit{RXTE} observations that were performed between 2008 August 4 and 2010 April 30 (under the Program IDs: P93019, P94019 and P05019) which cover the investigated LAT observing span. Individual \textit{RXTE} pointings are typically between 3 and 4 ks long and the total exposure time of all selected observations is about 196 ks. For each observation, we extracted events in the 2$-$10 keV range collected with the PCA and converted their arrival times to the solar system barycenter using the faxbary tool of HEASoft 6.8.

 
\section{Results}

\subsection{Search for Persistent Emission}

As evident from Figure \ref{fig:countmap}, \xsrc is clearly not detected in the LAT energy passband. A point source search using the 
filtered event list with the gtfindsrc tool of ScienceTools results in a potential source whose coordinates are 
inconsistent with that of \xsrcnos, therefore yields no detection. 

In order to obtain very-high-energy gamma-ray flux upper limits of \xsrcnos, we fitted the data from the $15^{\circ}$ radius 
region. We added all 
bright cataloged sources and recently discovered blazar \citep{vanb10} within this region of interest into the model 
as well as the galactic 
diffuse and extragalactic diffuse emission leaving their model parameters free. 
The fit yields a test statistics (TS) value of $\sim$0.23 
which implies a detection significance less than 1$\sigma$. The 3$\sigma$ flux 
upper limits with a power law index 2.5 are 2.32 $\times$ 10$^{-6}$ MeV cm$^{-2}$ s$^{-1}$ in the 0.2-1.0 GeV band 
and 1.28 $\times$ 10$^{-6}$ MeV cm$^{-2}$~s$^{-1}$ in the 1.0-10.0 GeV band. Note here that the 
spectral parameters and fluxes of the cataloged sources obtained in the latter fit are consistent with the 
catalog values\footnotemark \footnotetext{http://fermi.gsfc.nasa.gov/ssc/data/access/lat/1yr$\mathunderscore$catalog}, 
showing that our analysis is robust.

We performed similar spectral modelling for the data of the $2^{\circ}$ radius region with a power law model of index 3 as well as the galactic diffuse and extragalactic 
isotropic diffuse emission models. The resulting TS value is $\sim$3 which implies a detection significance less than 2$\sigma$. 
We chose the 0.2-1.0 and 1.0-10.0 GeV energy bands for flux calculations and find 3$\sigma$ upper limits to the source flux in these energy bands as 5.72 $\times$ 10$^{-6}$ MeV cm$^{-2}$ s$^{-1}$ and 1.29 $\times$ 10$^{-6}$ MeV cm$^{-2}$ s$^{-1}$, respectively.
In Figure \ref{fig:nufnu}, we present the high-energy gamma-ray flux upper limits of \xsrc in the $\nu$F$_{\nu}$ representation along with its low energy gamma-ray behavior \citep[the data obtained from]{dH08}. We discuss their implications in \S 4.

\subsection{Search for Pulsed Emission}

\begin{figure*}
\begin{center}
\includegraphics[width=0.8\textwidth]{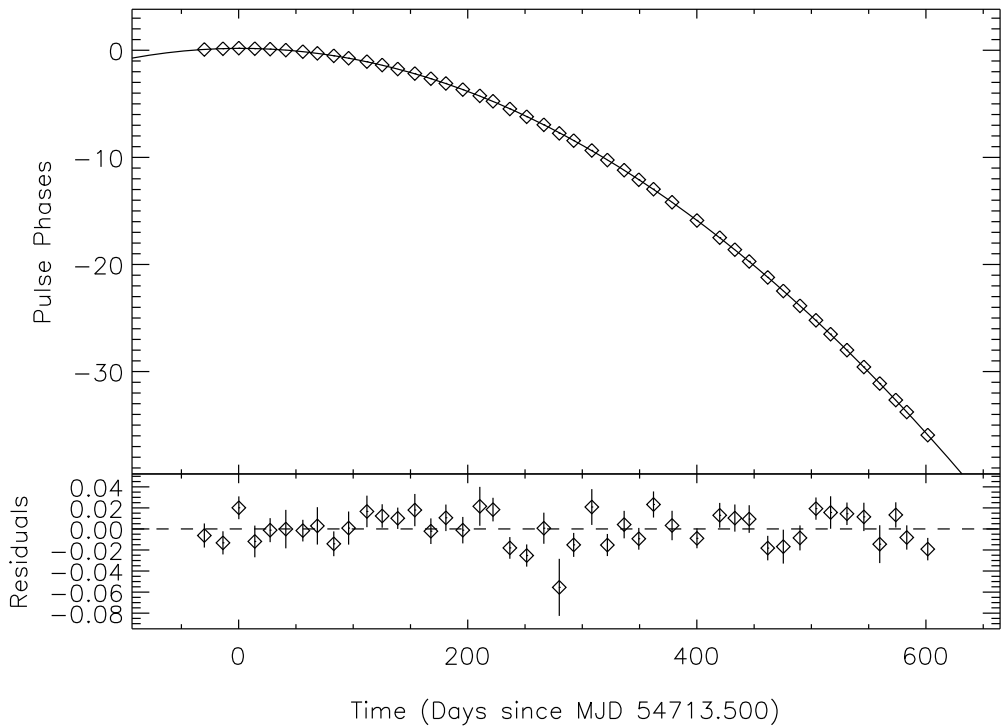}
\label{fig:spinphase}
\end{center}
\caption{(Top) spin phase shifts of \textit{RXTE}/PCA observations of \xsrc with respect to the epoch. 
The solid line is the best fitting model, that is a third-order polynomial. (Bottom) residuals of the fit.}
\end{figure*}

We employed a Fourier based epoch folding technique to obtain the spin ephemeris of \xsrc using \textit{RXTE}/PCA observations covering 
the time span of the LAT exposure of the source. We first generated the pulse profile of the source using three consecutive PCA 
observations around the epoch (MJD 54713.5). Then, we grouped observations in order for them to be spaced at least 0.2 days apart from 
each other, and we cross correlated the pulse profile of each group of pointings with the template profile to determine the phase 
shift of each pointing with respect to the template. Finally, we fit the phase shifts with a polynomial to obtain the spin ephemeris. 
In Figure \ref{fig:spinphase}, we present the phase shift and the best fitting model, that is a third-order polynomial 
($\chi$$^2$/degrees of freedom = 59.6/42). We tabulate the best fit spin ephemeris parameters of \xsrc in Table \ref{tab:poly_fit}.

\begin{table}[h]
\caption{Spin ephemeris of \xsrc}
\centering
\begin{tabular}{lc}\hline\hline
Parameter                   & Value   \\ \hline
Range (MJD)  &  54682.6 $-$ 55315.1 \\
Epoch (MJD)  	  & 54713.5  \\
$\nu$ (Hz)        & 0.1150900026(9) \\
$\dot{\nu}$ ($10^{-14}$ Hz s$^{-1}$)  &  $-$2.745(8) \\
$\ddot{\nu}$ ($10^{-23}$ Hz s$^{-2}$) &   3.6(3) \\ \hline
\end{tabular}
\label{tab:poly_fit}
\end{table}

To search for pulsed high-energy gamma-ray emission from \xsrcnos, we generated the LAT pulse profiles 
of the source in the 0.2-1.0 GeV and 1.0-10.0 GeV energy ranges using the precise PCA spin 
ephemeris obtained. We find that both LAT profiles are consistent with random fluctuations with respect to its mean. 
We calculate a 3$\sigma$ upper limit to the rms pulsed fraction of 1.5\% in the 0.2-1.0 GeV band and 2.3\% in the 1.0-10.0 GeV band. 
We also investigated the lower energy part of the LAT passband (30-200 MeV), which also resulted with no evidence of pulsed emission; 
the 3$\sigma$ rms pulsed fraction upper limit is 1.6\%.


\section{Discussion}

We searched for both persistent and pulsed high-energy emission from the AXP 4U 0142+61 using \textit{Fermi}/LAT data. We find no significant detection 
in either of the two objectives. Nevertheless, we obtained 3$\sigma$ upper limits for the high energy persistent emission in the 0.2-1.0 GeV 
and 1.0-10.0 GeV ranges of 5.72 $\times$ 10$^{-6}$ MeV cm$^{-2}$ s$^{-1}$ and 1.29 $\times$ 10$^{-6}$ MeV cm$^{-2}$ s$^{-1}$, respectively. 
As for the pulsed emission, a 3$\sigma$ upper limit to the RMS pulsed amplitude in the 0.2-1.0 GeV range is 1.5\% 
and in the 1.0-10.0 GeV range is 2.3\%. The search in the lower energy LAT passband (30-200 MeV) also did not yield a pulsed emission.
The 3$\sigma$ rms pulsed amplitude upper limit is 1.6\%. Our LAT upper limits are much lower than high pulsed fraction (up to 100\%) 
seen in hard X-rays with \textit{\igl} \citep{dH08}.

\begin{figure*}
\begin{center}
\includegraphics[width=0.8\textwidth]{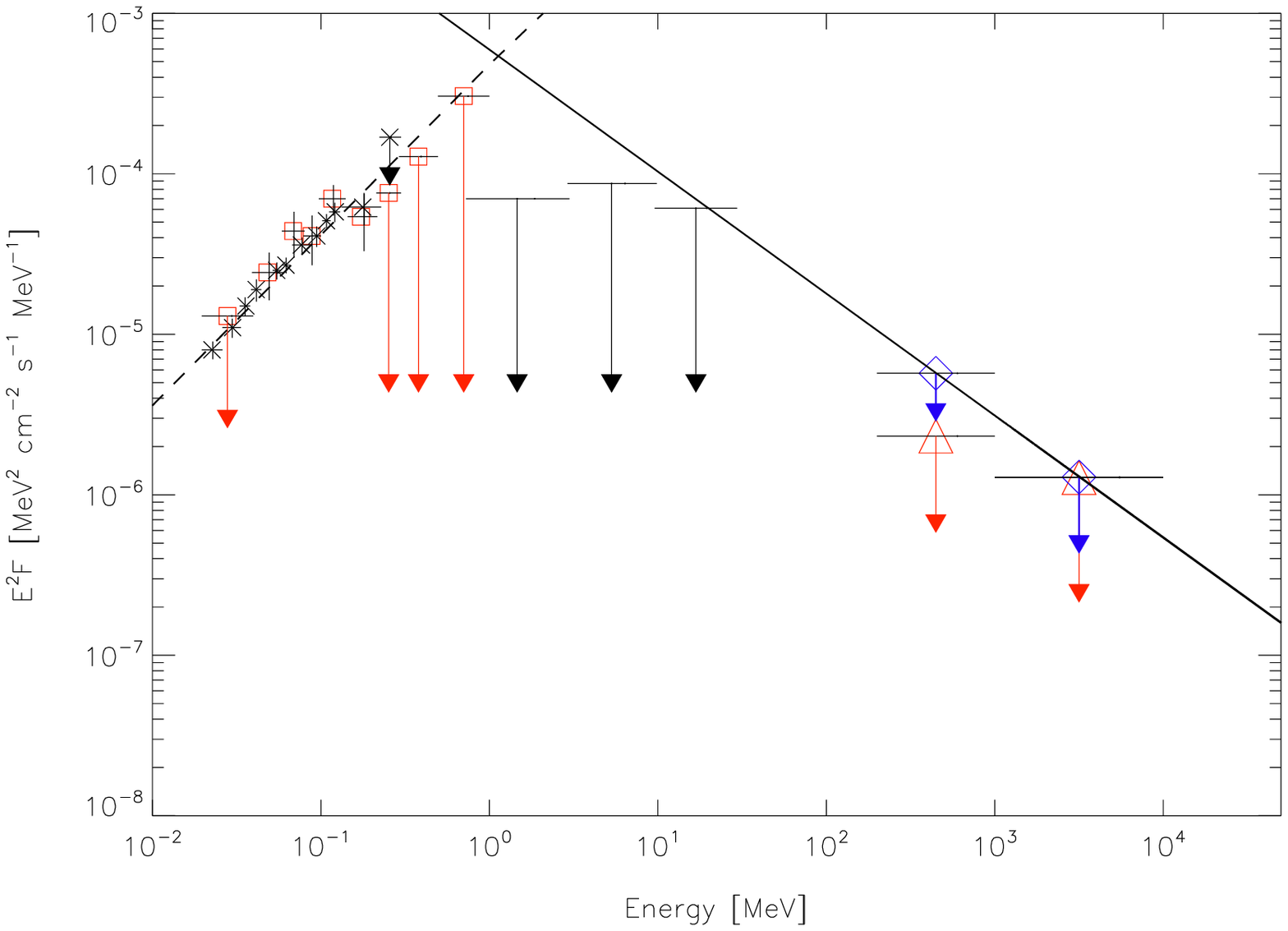}
\label{fig:nufnu}
\end{center}
\caption{Wide band $\nu$F$_{\nu}$ spectrum of 4U 0142+61: \textit{\iglnos}/ISGRI (20-300 keV) in black (stars), 
\textit{\iglnos}/SPI (20-1000 keV) in red (open squares) and \textit{CGRO}/COMPTEL (0.75-30 MeV) 2$\sigma$ upper limits in black 
(data constructed from \citet{dH08}). \textit{Fermi}/LAT upper limits (in the 0.2-1.0 GeV and 1.0-10.0 GeV) 
obtained using the $2^{\circ}$ extraction region are shown in blue diamonds and that of $15^{\circ}$ extraction region 
in red triangles. Dashed line is the best fit power law model to the ISGRI data points \citep{dH08}. 
Solid line shows the power law upper limit trend of the $2^{\circ}$ \textit{Fermi}/LAT region.}
\end{figure*}

In order to establish the spectral shape of \xsrc on a wide energy range, we constructed the $\nu$F$_{\nu}$ 
spectrum of the source in the 15 keV$-$10 GeV range by adopting the hard X-ray spectrum presented in \citet{dH08}
and placing the upper limits calculated in this work. \citet{dH08} fitted the \textit{\igl}/ISGRI 
data with a simple power law model of index 0.93 $\pm$ 0.06. We present this fit with dashed lines in Figure \ref{fig:nufnu}. 
We place an upper limit curve, which is the line connecting the two LAT upper limit measurements
(as shown with solid line in Figure \ref{fig:nufnu}), is also a power law with an index of $-$0.76. These two curves intersect with 
each other at $\sim$1.1 MeV which is an upper limit to the spectral break energy. 
Note that the spectral break upper limit is consistent with \citet{dH08} measurement of $279_{-41}^{+65}$ keV 
obtained by fitting a logparabolic function to the combined \textit{XMM-Newton}, \textit{\iglnos}/ISGRI, \textit{\iglnos}/SPI and \textit{CGRO}/COMPTEL data.

Our estimated upper limit to the spectral break energy is in accordance with the coronal emission model by 
\citet{bt07}. According to their model, photons with energies in excess of $\gtrsim$1 MeV would be trapped 
in the ultrastrong magnetic fields (\textit{B} $\gtrsim$ 10$^{14}$ G). In such a case photons would either split into 
two photons or they would create an electron-positron pair, therefore, suppressing the emission from the inner corona 
above $\sim$1 MeV. Our estimate of the spectral break energy upper limit also agrees with the predictions of the 
quantum electrodynamics model by \citet{hh05a, hh05b} as they expect the lower limit to the break energy to be around 1 MeV. 
On the other hand, they claim that if a source has a significant excess emission in optical wavelengths, as in the case of 4U 0142+61 
\citep{hulleman00}, its $\nu$F$_{\nu}$ spectrum should continuously increase in the 10-200 MeV range. 
Note that optical emission originates from the neutron star itself \citep{km02, dhillion05} and from a disk around 4U 0142+61 
\citep{wang06, es06}. However, the origin of the excess optical emission in 4U 0142+61 is not clear. If the excess optical 
emission originates from the compact object, our results place constraint, although not stringent, on the quantum electrodynamics 
model due to the lack of increase in the 10-200 MeV 
range in the $\nu$F$_{\nu}$ spectrum. On the other hand, if the disk provides a significant contribution to the observed emission, 
then the optical radiation from the neutron star itself would not be excessive and the quantum electrodynamics model would still remain feasible.

\acknowledgments 

We thank M. Ali Alpar and Yuki Kaneko for valuable discussions. S.\c{S}.M. and E.G. acknowledge EU FP6 Transfer of Knowledge Project “Astrophysics of Neutron Stars” (MTKD-CT-2006-042722).


\begin{thebibliography}{}

\bibitem[Atwood  et. al.(2009)]{atwood09} Atwood, W. B., et. al. 2009, ApJ, 697, 1071

\bibitem[Baring \& Harding(2007)]{bh07} Baring M. G., \& Harding, A. K. 2007, Ap\&SS, 308, 109

\bibitem[Beloborodov \& Thompson(2007)]{bt07} Beloborodov, A. M. \& Thompson, C. 2007, ApJ, 657, 967

\bibitem[den Hartog et al.(2004)]{dH04} den Hartog, R. P., Kuiper, L., Hermsen, W., \& Vink, J. 2004, ATel, 293, 1

\bibitem[den Hartog et al.(2008)]{dH08} den Hartog, R. P., Kuiper, L., Hermsen, W., Kaspi, V. M., Dib, R., Kn\"{o}dlseder, J., \& Gavriil, F. P. 2008, A \& A, 489, 245

\bibitem[Dhillon et al.(2005)]{dhillion05} Dhillion, V. S., Marsh, T. R., Hulleman, F., van Kerkwijk, M. H., Shearer, A.,  Littlefair, S. P., Gavrill, F. P., \&
Kaspi, V. M. 2005, MNRAS, 363, 609

\bibitem[Duncan \& Thompson(1992)]{dt92} Duncan, R. C., \& Thompson C. 1992, ApJ, 392, L9

\bibitem[Ertan \& \c{C}al\i\c{s}kan(2006)]{es06} Ertan, \"U., \& \c{C}al\i\c{s}kan, S. 2006, ApJ, 649, L87 

\bibitem[Fahlman \& Gregory(1981)]{fg81} Fahlman, G. G., \& Gregory, P. C., 1981, Nature, 293, 202

\bibitem[Heyl \& Hernquist(2005a)]{hh05a} Heyl, J. S., \& Hernquist, L. 2005a, ApJ, 618, 463

\bibitem[Heyl \& Hernquist(2005b)]{hh05b} Heyl, J. S., \& Hernquist, L. 2005b, MNRAS, 362, 777

\bibitem[Hulleman et al.(2000)]{hulleman00} Hulleman, F., van Kerkwijk, M.H., Kulkarni, S.R., 2000, Nature, 408, 689

\bibitem[Kern \& Martin(2002)]{km02} Kern, B., \& Martin, C. 2002, Nature, 417, 527

\bibitem[Kuiper et al.(2004)]{kuiper04} Kuiper, L., Hermsen, W., Mendez, M. 2004, ApJ, 613, 1173

\bibitem[Kuiper et al.(2006)]{kuiper06} Kuiper, L., Hermsen, W., den Hartog, P. R. 2006, ApJ, 645, 556

\bibitem[Mereghetti(2008)]{mereghetti08} Mereghetti, S., 2008, A\&AR, 15,225

\bibitem[Revnivtsev et al.(2004)]{revn04} Revnivtsev, M. G., et al. 2004, Astron. Lett., 30, 382

\bibitem[Thompson \& Duncan(1996)]{td96} Thompson, C., \& Duncan, R. C. 1996, ApJ, 473, 322

\bibitem[Vandenbroucke et al.(2010)]{vanb10} Vandenbroucke, J. et al. 2010, ApJ, 718, L166

\bibitem[Wang et al.(2006)]{wang06} Wang, Z., Kuiper, L., Hermsen, W., den Hartog, P. R. 2006, Nature, 440, 772

\bibitem[Woods \& Thompson(2006)]{wt06} Woods, P.M., \& Thompson, C. 2006, in Compact Stellar X-ray Sources, eds. W.H.G. Lewin \& M. van der Klis, Cambridge Univ. Press


\end{thebibliography}
\end{document}